\documentclass[preprint2]{aastex}

\shorttitle{Search for VHE gamma-ray emission from HBLs}
\shortauthors{J.~Albert et al.}

\begin{document}

\title{Systematic search for VHE gamma-ray emission from X-ray bright high-frequency BL Lac objects}
%
%
\author{
 J.~Albert\altaffilmark{a}, 
 E.~Aliu\altaffilmark{b}, 
 H.~Anderhub\altaffilmark{c}, 
 P.~Antoranz\altaffilmark{d}, 
 C.~Baixeras\altaffilmark{e}, 
 J.~A.~Barrio\altaffilmark{d},
 H.~Bartko\altaffilmark{f}, 
 D.~Bastieri\altaffilmark{g}, 
 J.~K.~Becker\altaffilmark{h},   
 W.~Bednarek\altaffilmark{i}
% ...
}
 \altaffiltext{a} {Universit\"at W\"urzburg, D-97074 W\"urzburg, Germany}
 \altaffiltext{b} {IFAE, Edifici Cn., E-08193 Bellaterra (Barcelona), Spain}
 \altaffiltext{c} {ETH Zurich, CH-8093 Switzerland}
 \altaffiltext{d} {Universidad Complutense, E-28040 Madrid, Spain}
 \altaffiltext{e} {Universitat Aut\`onoma de Barcelona, E-08193 Bellaterra, Spain}
 \altaffiltext{f} {Max-Planck-Institut f\"ur Physik, D-80805 M\"unchen, Germany}
 \altaffiltext{g} {Universit\`a di Padova and INFN, I-35131 Padova, Italy}  
 \altaffiltext{h} {Universit\"at Dortmund, D-44227 Dortmund, Germany}
 \altaffiltext{i} {University of \L\'od\'z, PL-90236 Lodz, Poland} 
% ...

\date{Received date / Accepted date}

\begin{abstract}

All but three (M87, BL Lac and 3C 279) extragalactic sources detected so far at very high energy (VHE) $\gamma$-rays belong to the class of high-frequency peaked BL Lac (HBL) objects. This suggested to us a systematic scan of candidate sources with the MAGIC telescope, based on the compilation of X-ray blazars by Donato et al.\ (2001).
The observations took place from December 2004 to March 2006 and cover sources on the northern sky visible under small zenith distances $zd <
30\degr$ at culmination, constraining the declination to values between $-2\degr$ and $+58\degr$. 
The sensitivity of the search was planned for detecting X-ray bright ($F(1\,\rm{keV}) > 2\,\mu\rm{Jy}$) sources emitting at least the
same energy flux at 200\,GeV as at 1\,keV. In order to avoid strong $\gamma$-ray attenuation close to the energy threshold, the redshift of the sources was constrained to values $z<0.3$.
Of the fourteen sources observed, 1ES 1218+304 (for the first time at very high energies) and 1ES 2344+514 
(strong detection in a low flux state) have been detected in addition to the known bright TeV blazars Mrk 421 and Mrk 501.  A marginal excess of 3.5$\sigma$ from the position of 1ES 1011+496 was observed and has been confirmed as a source of VHE $\gamma$-rays by a second MAGIC observation triggered by a high optical state \citep{alb2007e}. For the remaining sources, we present here the 99\% confidence level upper limits (u.l.) on the integral flux above $\sim$\,200\,GeV. 
We characterize the sample of HBLs (including all HBLs detected at VHE so far) by looking for correlations between their 
multi-frequency spectral indices determined from simultaneous optical, archival X-ray, and radio luminosities, finding that the VHE emitting
HBLs do not seem to constitute a unique subclass. The absorption corrected $\gamma$-ray luminosities at 200\,GeV of the HBLs are generally not higher than their X-ray luminosities at 1\,keV.
\end{abstract}


\section{Introduction}

Blazars belong to the most extreme objects in astronomy. Dominated by a non-thermal continuum spectrum, covering up to twenty decades in energy, they show  variability on time scales of years down to minutes \citep{alb2007f, aha2007a},  
and apparent luminosities exceeding $10^{49}\,\rm{erg}\,\rm{s}^{-1}$. 
Morphologically, blazars show strongly collimated jets extending from scales not much larger than the event horizon of a supermassive black hole \citep{biretta2002} up to megaparsec scales.  Superluminal motion of knots in the radio jets indicates relativistic bulk motion \citep{ghisellini1993}. 
X-ray knots at distances of more than the radiative cooling length from the nucleus indicate in situ particle acceleration \citep{biretta1991}, occuring at traveling and stationary shocks in the jet.
According to the unified scheme \citep[e.g.][]{urrypad95}, blazars are accreting supermassive black holes expelling a relativistic plasma jet at a small angle between the jet axis and the line of sight, with strong boosting of the observed emission due to relativistic bulk motion of the emitting plasma. 
BL Lacertae objects differ from the generally more luminous quasars 
by showing only faint or even absent emission lines, the absence of thermal big blue bump emission, and by not showing the otherwise typical luminosity evolution.

The Spectral Energy Distribution (SED) of blazars shows two pronounced peaks, the first between IR and hard X-rays, which is commonly believed to be synchrotron radiation of highly relativistic electrons, and the second one at $\gamma$-rays. Depending on the location of the first peak, BL Lac objects are further divided into low-frequency peaked BL Lacs (LBL, IR to optical) and high-frequency peaked BL Lacs (HBL, UV to X-rays) \citep{giommi1994}.
The second peak at high energies can be explained by inverse Compton scattering of low energy photons, produced as synchrotron radiation by the same population of electrons (Synchrotron Self Compton, SSC, \cite{mar}), or from ambient thermal photon fields, which could enter directly into the emission region \citep{der} or by scattering on material surrounding the jet \citep{sikora94}. The origin could also be due to hadronic processes associated
with proton and ion acceleration which leads to electromagnetic cascades and proton synchrotron radiation \citep{man,aha2000, pro}.
We cautiously remark, that the SED is probed at a sufficient level of sensitivity
only in a limited range, there are still large gaps, in particular between 50 keV and 100 MeV where
further peaks could show up.

In December 2004, when the regular observations with the Major Atmospheric Gamma-ray Imaging Cherenkov (MAGIC) telescope started, the  number of known VHE blazars was six, all of them X-ray bright HBL objects. At the time of writing, the number has increased to 19\footnote{For an up-to-date list of VHE blazars, see http://www.mppmu.mpg.de/~rwagner/sources/}, including one LBL object \citep[BL Lacertae,][]{alb2007d}, one Flat Spectrum Radio Quasar \citep[3C 279,][]{teshima2007} and the giant radio galaxy M87 \citep{aha2006b}. 

The detection of VHE $\gamma$-rays from cosmological distances is made difficult, due to absorption of $\gamma$-rays by photon-photon interactions with low energy photons from the evolving metagalactic radiation field (MRF). In the 100 GeV to 10 TeV range, far-infrared to optical photons are most important for the attenuation.  It has been realized that this leads to a relation between the $\gamma$-ray cutoff energy and
the source redshift known as the Fazio-Stecker relation \citep{fazio,kneiske04}.  
The fact that the so-far detected VHE sources have much lower redshifts compared with the EGRET 
$\gamma$-ray sources is in line with the expected effect of $\gamma$-ray attenuation, although 
the lack of curvature in the observed spectra is a source of serious doubts \citep{aha2006a}.

Due to the small field-of-view of an Imaging Air Cherenkov Telescope (IACT) and the limited duty cycle of $\sim$\,1000\,hrs per year, promising canditates for VHE emission have to be selected carefully. 
All established TeV sources are bright X-ray sources, most of them with comparable luminosities in both regimes, which renders a systematic scan of the X-ray brightest HBL objects a reasonable approach. 

Here we report on the results of such an approach pursued with the MAGIC telescope for a sample of 14 HBLs.
In Sect.\,2 the selection criteria for the sample are discussed, while the description of the observations can be found in Sect.\,3. The data analysis technique is described in Sect.\,4, and the analysis results are summarized in Sect.\,5.  
A brief explanation of how the SEDs were obtained from the data (using archival radio and X-ray data as well as simultaneous optical data)
can be found in Sect.\,6. Finally, the resulting 
properties of
the SED of HBLs and inferences on their luminosity function are discussed in Sect.\,7.

\section{HBL sample}
We used the compilation from \cite{don}, which provides 421 X-ray fluxes with spectral information for 268 blazars (136 of them HBL objects) together with average radio (at 5\,GHz) and optical (V-band) fluxes. The selection criteria were (i) redshift $z < 0.3$, (ii) X-ray flux $F_x(1\,\rm{keV}) > 2\,\mu\rm{Jy}$, and (iii) zenith distance ($zd < 30\degr$ during culmination, a total of fifteen sources remained after cuts.

The selection was made to avoid strong $\gamma$-ray attenuation at the energy threshold. At $z=0.3$, the expected cut-off energy is still above 200\,GeV, where MAGIC has its highest sensitivity. As the energy threshold increases with the zenith distance, all observations were 
carried out below $40\degr$, where the analysis energy threshold is around 200\,GeV.
As most of the established TeV sources show comparable luminosities in X-rays and in $\gamma$-rays, only the X-ray brightest HBLs were selected, leading to a cut at $2\,\mu\rm{Jy}$. Assuming the same luminosity at $\sim$\,200\,GeV, it corresponds to $\sim$\,7\% of the flux of the Crab Nebula, which would be detectable for MAGIC within 15\,hrs.
 
The goal was to observe them for at least 15\,hrs in order
to establish new VHE sources and to put constraints on the
SED of HBLs in a systematic fashion. 
The complete set is listed in Table\,\ref{sample}.

\begin{deluxetable}{llcccclc}
\tablewidth{0pt}
\tablecaption{List of targets.
\label{sample}} 
\tablehead{
\colhead{Source} &
\colhead{RA} &
\colhead{dec} &
\colhead{$z$} &
\colhead{flux \tablenotemark{a}} &
\colhead{$\Gamma$\tablenotemark{b}} &
\colhead{season} &
\colhead{Time [h]\tablenotemark{c}}}
\startdata
1ES 0120+340 &  01 23 08.9  & +34 20 50 & 0.272 & 4.34 & 1.93 & 2005 Aug-Sep &  14.9\\ 
RX J0319.8+1845 &  03 19 51.8 & +18 45 35 & 0.190 & 1.76\tablenotemark{d} & 2.07 & 2004 Dec-2005 Feb & 6.9\\ 
 &   &  &  & &  & 2005 Sep-2006 Jan & 4.7\\ 
1ES 0323+022  &  03 26 14.0  & +02 25 15 & 0.147 & 3.24 & 2.46 & 2005 Sep-Dec & 11.4\\ 
1ES 0414+009 &  04 16 53    & +01 04 54 & 0.287 & 5.00 & 2.49 & 2005 Dec-2006 Jan & 17.8 \\ 
1ES 0806+524   &  08 09 49.2 & +52 18 58 & 0.138 & 4.91 & 2.93 & 2005 Oct-Dec & 17.5\\ 
1ES 0927+500  &  09 30 37.6 & +49 50 24 & 0.188 & 4.00 & 1.88 & 2005 Dec-2006 Mar & 16.1\\
1ES 1011+496  &  10 15 04.2  &  +49 26 01 & 0.212\tablenotemark{e} & 2.15 & 2.49 & 2006 Mar-Apr & 14.5\\
Mrk 421       &   11 04 27.3 & +38 12 31.8 & 0.030 & 39.4 & 2.96 & 2004 Nov-2005 Mar & 25.6\tablenotemark{f}\\ 
1ES 1218+304 &  12 21 21.9   & +30 10 37 & 0.182 & 8.78 & 2.34 & 2005 Jan & 8.2\tablenotemark{g}\\ 
 &  &  &  &  &  & 2006 Jan-Mar & 14.6\\ 
RX J1417.9+2543  &  14 17 56.6  & +25 43 25 & 0.237 & 3.58  & 2.25 & 2005 Apr-Jun & 13.0\\ 
1ES 1426+428  &  14 28 32.5   & +42 40 25 &  0.129 & 7.63 & 2.09 & 2005 Mar-Dec & 6.1\\ 
Mrk 501        &  16 53 52.2 & +39 45 36.6 & 0.034 & 20.9 & 2.25 & 2005 May-Jul & 32.2\tablenotemark{h}\\ 
RX J1725.0+1152 &  17 25 04.4  & +11 52 16 & $>0.17$\tablenotemark{i} & 3.60 & 2.65 & 2005 Apr & 5.3 \\
1ES 1727+502 &  17 28 18.6  & +50 13 11 &  0.055 & 3.68 & 2.61 & -\tablenotemark{j} & 0\tablenotemark{j}\\ 
1ES 2344+514  &  23 47 04.9  & +51 42 18 &  0.044 & 4.98 & 2.18 & 2005 Aug-2006 Jan & 23.1\tablenotemark{k}\\ 
\enddata
\tablenotetext{a}{$F(1\,\rm{keV})\,[\mu\rm{Jy}]$, average value from different measurements, taken from \cite{don}.}
\tablenotetext{b}{$\Gamma$ is the spectral index for the differential spectrum (dN/dE) at 1\,keV, assuming a power law}
\tablenotetext{c}{Effective observation time after quality selection}
\tablenotetext{d}{Two measurements are above $3\,\mu$Jy, one below $1\,\mu$Jy}
\tablenotetext{e}{The earlier reported redshift of 0.200 was recently revised by \cite{alb2007e}}
\tablenotetext{f}{Results published in \cite{alb2007c}}
\tablenotetext{g}{Results published in \cite{alb2006b}}
\tablenotetext{h}{Results published in \cite{alb2007f}}
\tablenotetext{i}{The earlier reported redshift of 0.018 was recently revised by a lower limit \citep{sbaru2006}}
\tablenotetext{j}{Proposed but not observed due to bad weather}
\tablenotetext{k}{Results published in \cite{alb2007b}}
\end{deluxetable}

\section{Observations}
The MAGIC telescope is a single dish IACT, located on the Canary island of La Palma (28.8\degr\,N, 17.8\degr\,W, 2200\,m a.s.l.). A 17\,m diameter tessellated parabolic mirror with a total surface of $234\,\mbox{m}^2$, mounted on a light-weight space frame made from carbon fiber reinforced plastic tubes, focuses Cherenkov light from air showers, initiated by $\gamma$-rays or charged cosmic rays, onto a 576-pixel photomultiplier camera with a field-of-view of $3.5\degr$. The analogue signals are transported via optical fibers to the trigger electronics and each channel is read out by a 300 Msample/s FADC. Further details on the telescope can be found in \citet{MAGIC-commissioning} and \citet{CortinaICRC}. Note that the readout system has been upgraded to a 2GSample/s FADC in February 2007 \citep{goebel2007}. A second telescope of the same size for observations in stereo mode is currently under construction.

The observations took place from December 2004 to March 2006 in moonless nights. The data are taken in different observation modes. If the telescope is pointing to the source (on-mode), the background has to be determined by so-called off-data, where the telescope points to a nearby sky region where no $\gamma$-ray source is expected. The off-data cover the same $zd$ range with a similar night-sky background light intensity. The larger fraction of the source sample was observed in the so-called wobble mode, 
where the pointing of the telescope wobbles every 20 minutes between two symmetric sky locations with an angular distance of $0.4\degr$ to the source.
The background in the signal region can be estimated from sky locations placed at the same distance from the camera center as the candidate source.

Except 1ES 0927+500 and 1ES 0414+009, all objects were monitored by the KVA telescope (\textit{http://users.utu.fi/kani/1m/index.html}) on La Palma in the optical R-band. None of the sources showed flaring activity in the optical during the MAGIC observations. The host galaxy corrected fluxes \citep{nilsson}, taken simultaneously, averaged over the time of the MAGIC observations, are listed in Table\,\ref{discussion}.

\section{Data analysis technique}

The data are processed using the MAGIC Analysis and Reconstruction Software (MARS) \citep{mars}. A description of the different analysis steps can be found in \citet{gaug} (including the calibration) and \citet{bre2}.

As the trigger rate strongly depends on the weather condition, only data with a rate above 160\,Hz are used to ensure a high data quality. 

The moments up to third order of the light distribution are used to characterise each event by a set of image parameters
\citep{Hillas_parameters}. For background suppression, a SIZE-dependent parabolic cut in WIDTH $\times$ LENGTH is applied \citep{dyn_cuts}. To reconstruct the origin of the shower in the camera plane, the DISP method is employed \citep{les} to estimate
the distance between the centre of gravity of the shower and its origin. The third moment determines the direction of the shower
development. The constant coefficient $\xi$ from the parametrisation of DISP in the original approach is replaced in this analysis by $\xi_0 +
\xi_1\cdot \left(\mbox{LEAKAGE}\right)^{\xi_2}$, LEAKAGE being the fraction of light contained in the outermost camera
pixels. Thereby the truncation of the shower images at the camera border is taken into account. These coefficients are determined separately for on-off and wobble data using simulated $\gamma$-showers, which are produced by CORSIKA, version 6.023 \citep{cor,maj} for $zd$ below $40\degr$ and energies between 10\,GeV and 30\,TeV, following a power law with a spectral index -2.6.

The cut coefficients for the background suppression are optimised using Crab Nebula data, taken at similar $zd$. One set of cut coefficients is derived for data taken in on-off mode and one for wobble mode. The significance of a possible signal is determined from the distribution of the squared angular distance ($\theta^2$) between the shower origin and the source position.
The signal region is determined as $\theta < 0.23\degr$, corresponding to slightly more than two times the $\gamma$ point-spread function of the MAGIC telescope. 

For observation in on-off mode, the off-data have to be scaled to match the on-data. This is done in the region $0.37\degr < \theta < 0.80\degr$, where no bias from the source is expected. For wobble observations three regions, located symmetrically on a ring around the camera centre with the same distance from the centre as the source position, are defined as background regions. The scale factor is fixed to 1/3.

For every source the statistical significance according to equation 17 from \cite{lima} is calculated. The upper limit on the excess rate is derived with a confidence level (c.l.) of 99\%, using the method from \cite{rol}, which takes also the scaling factor of the background into account. The upper limit for the excess rate is then compared to the excess rate of the Crab Nebula, which leads to an upper limit of the flux in units of the Crab Nebula flux above a certain energy threshold, assuming a Crab like spectrum. The energy threshold is here defined as the energy where the differential distribution (dN/dE vs.\ E) of simulated $\gamma$-ray events, surviving all cuts, peaks. Note that this threshold depends also on the spectral shape.

\begin{deluxetable}{lcccccccc}
\tablewidth{0pt}
\tablecaption{Observations of the Crab Nebula used for the upper limit calculation \label{Crab}} 
\tablehead{
\colhead{} &
\colhead{} &
\colhead{Exp.} &
\colhead{$zd$} &
\colhead{$E_{\rm{thres}}$}&
\colhead{Excess} &
\colhead{Background} &
\colhead{} & 
\colhead{Sign.}\\ 
\colhead{Season} &
\colhead{Mode} &
\colhead{[hrs]} &
\colhead{degr.} &
\colhead{GeV}&
\colhead{[$\rm{min}^{-1}$]} &
\colhead{[$\rm{min}^{-1}$]} &
\colhead{scale} & 
\colhead{$[\sigma/\sqrt{h}]$}} 
\startdata
2005 Oct - 2006 Mar & w  &  8.6 & 23.0 & 230 & 6.43 & 6.23 & 0.33 & 14.6\\
2004 Dec - 2006 Mar & on & 38.7 & 16.2 & 190 & 8.80 & 9.19 & 0.92 & 13.5\\
\enddata
\end{deluxetable}

\begin{figure}[h]
\plotone{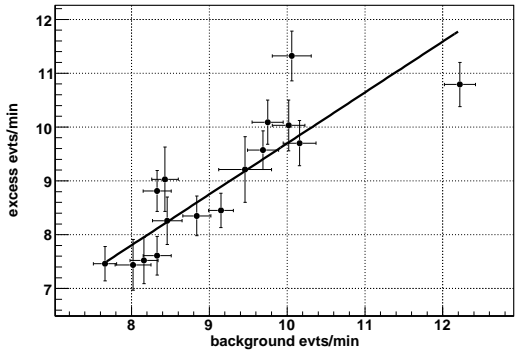} 
\caption{Excess rate ($excess$) vs.\ background rate ($bgd$) for the Crab Nebula (on-off mode). A linear fit yields $(0.95\pm0.10)\cdot bgd + 0.2 = excess$.\label{exon}}
\end{figure}

\begin{figure}[h]
\plotone{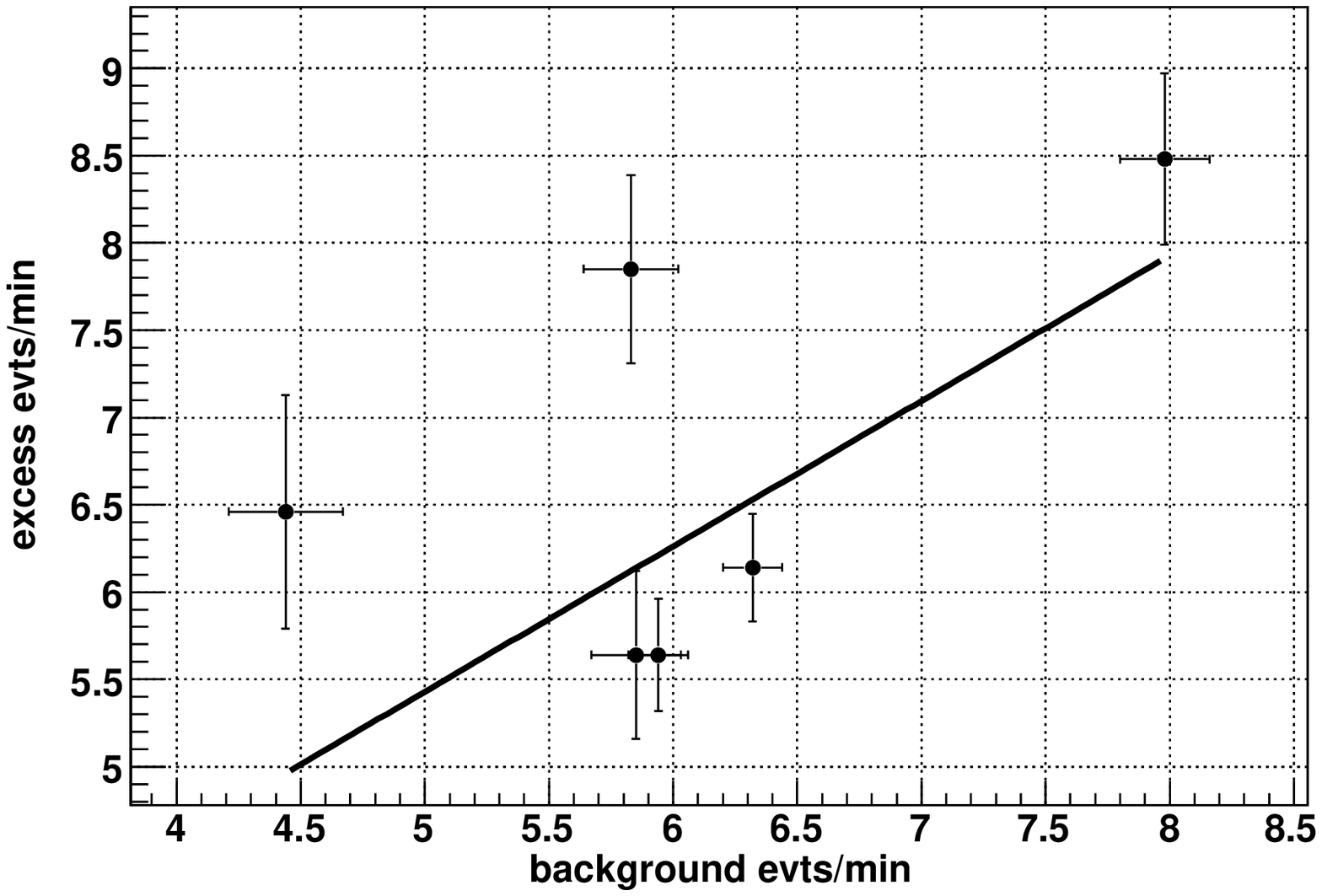} 
\caption{Excess rate ($excess$) vs. background rate ($bgd$) for the Crab Nebula (wobble mode). A linear fit yields $(1.26\pm0.3)\cdot bgd - 1.5 = excess$.\label{exwob}}
\end{figure}

A large sample of data from the Crab Nebula in on-off as well as in wobble mode is used, spread over the entire observation campaign (see Table\,\ref{Crab}). This analysis shows, that the excess rate of the Crab Nebula is correlated to the background rate (after $\gamma$-hadron separation). Therefore, depending on the background rate of the AGN, a reference value for the excess rate of Crab has to be calculated. This can be understood when taking into account that even after quality selection the rates fluctuate up to 20\%, depending on weather conditions. In Fig.\,\ref{exon} and Fig.\,\ref{exwob} the rate of excess events vs.\ the background rate of different subsamples of the Crab Nebula is shown. 
A linear regression to the on-off samples results in an acceptable fit ($\chi^2/\rm{d.o.f.} = 26/14$), showing a clear correlation. The fit for the wobble data is quite poor ($\chi^2/\rm{d.o.f.} = 13.5/3$). 
As a constant fit gives even a worse result and the fit on the on-data shows a correlation between background- and excess rate, the linear fit for the wobble data is also used to calculate the reference values for the comparison of the excess rates.
The Crab units are converted into a flux of  $\rm{photons}\,\rm{cm}^{-2}\,\rm{s}^{-1}$ using the spectrum of the Crab Nebula from \cite{alb2007g}.

The systematic error for the flux is estimated to be $\sim\,$30\% (see \cite{alb2007g} and discussion therein). For the u.l.\ determination there is also the uncertainty of the correct energy threshold (which depends on the source spectrum). 

\section{Results of the MAGIC observations}
Within this observation program, VHE $\gamma$-rays were discovered from 1ES 1218+304 \citep{alb2006b} and 1ES 2344+515 was observed in a low flux state with high significance \citep{alb2007b}. 
Mrk 421 was observed for more than 25\,hrs in 2005. The results are discussed in detail in \cite{alb2007c}. Mrk 501 was observed from May to July 2005 with more than 30\,hrs, revealing a high precision lightcurve on a day-by-day basis as well a two exceptionally short-time flares (see \cite{alb2007f} for more details). 
For ten sources of the sample, no significant signal is seen. The 2006 observation of 1ES 1218+304 results in a weak signal of $4.6\,\sigma$ (see Sect.\,5.2). A slightly refined analysis of 1ES 1011+496 yields a hint of a signal with a $3.5\,\sigma$ significance (see Sect.\,5.3). The results are listed in Table\,\ref{results}.  Observations of 1ES 1727+502 are still pending.

\subsection{Upper limits}
The u.l.s are between 2.3\% and 8.6\% of the Crab Nebula flux. For a Crab-like spectrum the energy thresholds vary between $(190 \pm 15)$\,GeV and $(230 \pm 15)$\,GeV, depending on the $zd$ of the observation. For the threshold calculation the exact $zd$ distribution of every observation is taken into account. 
As the Crab spectrum at $\sim$\,200\,GeV is quite hard (spectral slope -2.26 for the differential energy spectrum), the u.l.s are also calculated for an -3.0 power law spectrum, which represents quite well the average slope of all HBLs detected at VHE so far. In Table\,\ref{discussion} the energy threshold as well as the flux u.l.\ at 200\,GeV are given under the assumption of a -3.0 power law.

\begin{deluxetable}{lcccccccrcc}
\tablewidth{0pt}
\tablecaption{Results of the analysis.\label{results}}
\tablehead{
\colhead{} &
\colhead{} &
\colhead{Exp.} &
\colhead{$Zd$ \tablenotemark{a}} &
\colhead{$E_{\rm{thres}}$\tablenotemark{b}}&
\colhead{} &
\colhead{} &
\colhead{} &
\colhead{Sig.} &
\colhead{UL} &
\colhead{UL} \\
\colhead{Source} &
\colhead{Mode} &
\colhead{[hrs]} &
\colhead{[degr.]} &
\colhead{[GeV]} &
\colhead{Excess\tablenotemark{c}}&
\colhead{Backgr.} &
\colhead{Scale} &
\colhead{$\sigma$} &
\colhead{c.u.\tablenotemark{d}} &
\colhead{f.u.\tablenotemark{e}}}
\tablecolumns{11}
\startdata
1ES 0120+340 & w & 14.9 & 12.2       & 190 & -48 & 5358 & 0.33 & -0.6 & 0.032 & 0.75\\
RX J0319.8+1845 & w & 4.7 & 14.3    & 190 &  9   & 2225 & 0.33  & 0.2  & 0.049 & 1.15\\
RX J0319.8+1845 & on & 6.5 & 14.2   & 190 & -95 & 3257& 0.86  & -1.2 & 0.033 & 0.78\\
1ES 0323+022  & w & 11.4   & 29.0    & 230 & 55  & 5262 & 0.33 & 0.7 &  0.064 & 1.16\\
1ES 0414+009 &  w & 17.8 & 29.7     & 230 & 176 & 7309 & 0.33 & 1.8 & 0.057 & 1.03\\
1ES 0806+524 & w & 17.5 & 26.8      & 230 & 111 & 6174 & 0.33 & 1.2 & 0.056 & 1.01 \\
1ES 0927+500 & w & 16.1 & 22.1       & 230 & 72  & 5721 & 0.33 & 0.8 & 0.052 & 0.94 \\
1ES 1011+496 & w & 14.5 & 23.6       & 230 & 200 & 4857 & 0.33 & 2.5 & 0.086 & 1.55 \\
1ES 1218+304 & w & 14.6 & 26.6       & 230 & 400 & 5423 & 0.33 & 4.6 & 0.073\tablenotemark{f} & 1.31\tablenotemark{f} \\
RX J1417.9+2543 & on & 13.0 & 9.7   & 190 & -137& 9007 & 1.03 & -1.0 & 0.023 & 0.54\\
1ES 1426+428  & on & 6.1 & 16.6      & 190 &  -7   & 2561 & 0.24 & -0.1 & 0.050 & 1.18\\
RX J1725.0+1152 & on & 5.3 & 17.4  & 190  & -69 & 2001 & 0.98 & -1.1 & 0.046 & 1.08\\
\enddata
\tablenotetext{a}{Mean zenith angle of the observation}
\tablenotetext{b}{Peak response energy for a Crab like spectrum}
\tablenotetext{c}{Background subtracted signal events for $\theta<0.23\degr$}
\tablenotetext{d}{Integral flux above the threshold given in units of the flux of the Crab Nebula (crab units, c.u.)}
\tablenotetext{e}{Integral flux above the threshold given in flux units (f.u.) of $10^{-11}\,\rm{photons}\,\rm{cm}^{-2}\,\rm{s}^{-1}$}
\tablenotetext{f}{Integral flux above threshold in cu.\ and f.u.\ corresponding to the $4.6\,\sigma$ excess}
\end{deluxetable}

\subsection{1ES 1218+304}
\begin{figure}[h]
\plotone{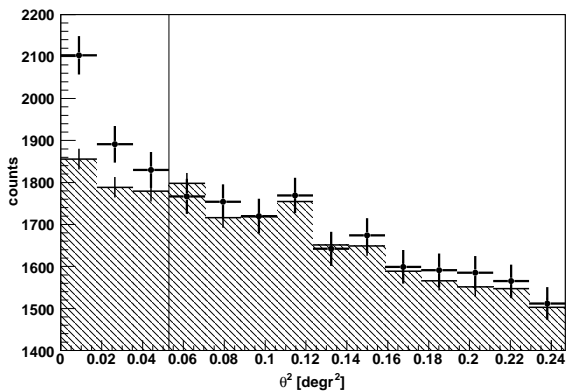} 
\caption{Distribution of the squared angular distance with respect to the position of 1ES\,1218+304 (dots) and to three off-regions (scaled by 1/3, gray shaded area). \label{theta}}
\end{figure}

The source was observed in 2006 from January the 29th to March the 5th during 15 nights with in total 14.6 hrs. 
Figure\,\ref{theta} shows the distribution of the squared angular distance between the reconstructed shower origin of each event and the assumed source position. The vertical line indicates the signal region. The background is determined by three off-regions in the camera. The excess has a statistical significance of $4.6\,\sigma$ (see also Table\,\ref{results}).

Under the assumption of a power law spectrum with spectral slope of -3.0 as measured in 2005 \citep{alb2006b}, the energy threshold decreases with respect to a Crab like spectrum to 190\,GeV. The average integral flux above 180\,GeV for the complete sample is $F( > 180\,\rm{GeV}) = (1.48\pm0.48) \times 10^{-11}\,\rm{photons}\,\rm{cm}^{-2}\,\rm{s}^{-1}$.

The search for flux variability on time scales from days ($\sim$\,1\,h observation time per night) to weeks (sample with several consecutive nights) yielded no significant variability.

The integral flux above 180\,GeV indicates a $\sim$\,30\% lower flux than the one measured in 2005, even though they are consistent within their errors. The average optical flux in 2006 was $\sim$\,20\% lower than in 2005, which is already significant compared to the statistical error of $\sim$\,2\%. During the observations in 2006 the optical flux decreased continuously from $(1.144\pm0.036)\,\rm{mJy}$ on the 3rd of February to $(0.947\pm0.038)\,\rm{mJy}$ on the 7th of March (all optical fluxes are host galaxy subtracted). This trend continued until June, where the source remained in a low optical state. Unfortunately with six nights the sampling of the optical light curve during the MAGIC observations is quite low, hence an increase of the optical activity on a time scale of days can not be excluded.

\subsection{1ES 1011+496}
The standard analysis performed for the whole source sample yields a significance of $2.5\,\sigma$, which gives already a hint for a possible signal. In a more refined analysis, the cut in $\theta^2$, which determines the signal region, was reduced to $\theta = 0.20\degr$. In case of a weak signal the increased signal to background ratio would lead to a higher significance. Also the SIZE-dependent cut for the background suppression is changed to a slightly higher value.
The same analysis, performed on a data sample of the Crab Nebula, results in almost the same significance as with the standard coefficients, but with a $\sim$\,13\% lower $\gamma$-rate due to the reduced $\gamma$ acceptance and a 37\% lower background rate.

This analysis yields $3.5\,\sigma$ for 1ES 1011+496 which - if interpreted as a detection - corresponds to an integral flux of $F(>180\,\rm{GeV}) = (1.26\pm0.40)\times10^{-11} \rm{photons}\,\rm{cm}^{-2}\,\rm{s}^{-1}$. Further observations with the MAGIC telescope, triggered by an optical outburst in March 2007, shows a clear signal of $6.2\,\sigma$ within 18.7\,hrs of observation, resulting in flux $\sim\,50\%$ higher than in 2006 \citep{alb2007e}.

\subsection{1ES 1426+428}
The VERITAS collaboration reported a steep spectrum above 300\,GeV for their observations in 2001, well fitted by a power law with spectral index $-(3.50\pm0.35)$ \citep{pet}. Extrapolating the spectral fit to 200\,GeV, it yields an integral flux of 0.50 Crab above 200\,GeV, which is by a factor of 10 larger than the u.l.\ presented in this work. Previous measurements yield a marginal detection in 2000 and upper limits for the data taken from 1995 to 1999 \citep{hor} with the most stringent one of 0.08 Crab above 350\,GeV. 

The HEGRA collaboration published a much harder spectrum at higher energies (above $\sim$\,800\,GeV) for their combined 1999 and 2000 data, which were well fitted by a power law with spectral index $-(2.6\pm0.6)$ \citep{aha02}. An extrapolation of the power law yields an integral flux above 200\,GeV of 0.075 Crab. Due to the large extrapolated energy range, combined with the large statistical error of 0.6 for the slope, the uncertainty is a factor of two. Further measurements in 2002 with 
the HEGRA telescopes showed the source in a 2.5 times lower flux state \citep{aha03}. 

The u.l.\ for the flux above 200\,GeV presented in this work, indicates a lower flux than measured from 1999 to 2001 during several campaigns with different telescopes, whereas it is consistent with the low flux level observed in 2002.

\section{Spectral Energy Distribution}
The complete set of sources, described in Sect.\,2, amounts to 14 objects (without 1ES 1727+502) and includes the five established TeV sources Mrk 501, 1ES 1218+304, 1ES 1426+428, Mrk\,421 and 1ES 2344+514. To better understand the spectral energy distribution
of VHE gamma ray emitting blazars, the sample is enlarged by including ten more sources which fullfill the same selection
criteria in X-ray flux and redshift, but which deviate only in the zenith angle cut.  This means, they can only be observed under large zenith
distances from the MAGIC site or not at all, in which case there is still information available from corresponding H.E.S.S.\ observations.
An exception is PG 1553+113 where the redshift is not known. For the total, enlarged sample of 24 HBLs, multi-wavelength data are collected in the following bands: radio  (5\,GHz, \cite{don}),  optical (R-band, 640\,nm, simultaneous data from KVA or, if not available from \cite{don}), X-rays (1\,keV, \cite{don}), and $\gamma$-rays (200\,GeV, see Table\,\ref{rest}, Table\,\ref{discussion} and ref.\ in Table\,\ref{sample}). For 1ES 1426+428, in addition to the upper limit derived in this work, the extrapolation of the spectrum, as measured by HEGRA in 1999/2000, is used to mark the detected flux. 
The optical data were corrected for galactic extinction, using the coefficients from the NASA Extragalactic Database (NED), which are calculated according to \cite{schlegel}.

For the $\gamma$-ray flux, sizeable attenuation is expected from current models of the MRF \citep{dwek,kneiske04}. Therefore all u.l.\ at 200\,GeV as well as the measured fluxes of the detected HBLs are corrected for the absorption by multiplying with $\exp(\tau(200\,\rm{GeV}, z))$ where $\tau$ denotes the pair production optical depth. The "best-fit 2006" MRF-model (Kneiske 2007, in preparation) is employed. This model is based on the "best fit" model from \cite{kneiske04}, but with a lower star formation rate to keep the energy density in the optical band closer to the lower limits, derived from the galaxy number counts. It is also consistent with the u.l. derived by \cite{aha2006a} from the VHE spectrum of 1ES 1101-233 ($z=0.186$).
The optical depth values for 200\,GeV photons for all sources of this sample are listed in Table\,\ref{discussion}. 


All fluxes are K-corrected. Radio spectral indices\footnote{Spectral index defined by the photon flux $F \propto \nu^{-\alpha}$ [photons\,$\rm{cm}^{-2}\,\rm{s}^{-1}$]} of ten of the sources can be found in \cite{landt}. For the other 14 objects the average value $\alpha_{\rm{R}} = 0.23$ of the ten sources is used. For the optical data, the spectral indices of nine sources, calculated at slightly higher wavelengths, are taken from \cite{bersanelli}. For the other 15 objects, the average value $\alpha_{\rm{O}} = 0.65$ of the nine sources is used. At 1\,keV, the spectral indices are taken from \cite{don}, except for 1ES 0229+200, which is not included in this compilation. Instead, the flux is taken from \cite{CostGhis} together with the average value for the spectral index $\alpha_{\rm{X}} = 1.36$ of all other sources. At 200\,GeV the measured spectral indices are used for the detected sources, while for the non-detected ones the average value $\alpha_{\rm{\gamma}} = 2.0$ is used. To take into account the energy dependent attenuation at VHE, which causes a hardening of the spectrum, the measured spectral indices are changed by -0.4 for $0.1 < z < 0.2$, -0.8 for $0.2 < z < 0.3$ and remain unchanged for $z < 0.1$. 

A special treatment is necessary to derive the flux at 200\,GeV from 1ES 0229+200, recently discovered at VHE $\gamma$-rays \citep{aha2007c}. The spectrum is measured above 580\,GeV, well fitted by a power law with spectral index $-(2.51\pm0.19)$. As the source is located at $z=0.1396$, strong absorption is expected at these energies. Therefore the spectrum is first deabsorbed and afterwards extrapolated to lower energies (Table\,\ref{rest}). The resulting intrinsic spectrum is well described by a power law with a spectral index of $-(1.09\pm0.25)$ (flux normalisation: $(4.24\pm0.81)\times10^{-12}\,\rm{cm^{-2}\,s^{-1}\,TeV^{-1}}$ at 1\,TeV). This result is in good agreement with the results from \cite{stecker2007}, yielding model dependent intrinsic spectral indices in the range from $1.1\pm0.3$ to $1.5\pm0.3$.

After these corrections, the broad-band spectral indices\footnote{$\alpha_{1-2}= -\log(F_1/F_2)/\log(\nu_1/\nu_2) \; , \;\nu_1 < \nu_2$} $\alpha_{1-2}$ between the different energy regimes are calculated. Also the luminosities $\nu L_{\nu}$ are calculated assuming isotropic emission and with the use of the following cosmological parameters: $H_0 = 71\,\rm{km}\,\rm{s}^{-1}\,\rm{Mpc}^{-1}$, $\Omega_{\Lambda}=0.73$, and $\Omega_m = 0.27$.

\begin{deluxetable}{lcc}
\tablewidth{0pt}
\tablecaption{HBLs detected at VHE which do not belong to the sample described in Sect.\,2 .\label{rest}} 
\tablehead{
\colhead{Source} &
\colhead{$F_{\gamma}$\tablenotemark{a}} &
\colhead{Ref.}}
\startdata
1ES 0229+200 & 1.6\tablenotemark{b}   & \cite{aha2007c} \\
1ES 0347-121 & 4.25 & \cite{aha2007b} \\
PKS 0548-322 & 1.9  & \cite{superina07}\\
1ES 1101-232 & 2.93 & \cite{aha2006a}\\
Mrk 180 & 11.0 & \cite{alb2006c}\\
PG 1553+113   & 11.5 & \cite{alb2007a}\\
1ES 1959+650 & 17.4 & \cite{alb2006a}\\
PKS 2005-304 & 6.63 & \cite{aha2005a}\\
PKS 2155-489 & 26.3 & \cite{aha2005b}\\
H 2356-309  & 2.78 &  \cite{aha2006a}\\
\enddata
\tablenotetext{a}{$\nu F_{\nu}(200\,\rm{GeV})$ in units of $10^{-12}\,\rm{erg}\,\rm{cm}^{-2}\,\rm{s}^{-1}$}
\tablenotetext{b}{This value comes from an extrapolation as described in the text}
\end{deluxetable}

\begin{deluxetable}{lcccccrc}
\tablewidth{0pt}
\tablecaption{
Upper limits on the $\gamma$-ray flux at 200\,GeV under the assumption of a -3.0 power law spectrum togehter with the optical depth and the simultaneous optical data.
\label{discussion}} 
\tablehead{
\colhead{} &
\colhead{} &
\colhead{$E_{\rm{thres}}$\tablenotemark{a}} &
\colhead{} &
\colhead{} &
\colhead{} &
\colhead{} \\
\colhead{Source} &
\colhead{Mode} &
\colhead{GeV} &
\colhead{$F_{\gamma}$\tablenotemark{b}} &
\colhead{$z$} &
\colhead{$\tau$\tablenotemark{c}}&
\colhead{$F_{\rm{o}}$\tablenotemark{d}} &
\colhead{$F_{\rm{X}}$\tablenotemark{e}}}
\tablecolumns{7}
\startdata
1ES 0120+340    & w   & 170   & 4.0 & 0.272    & 0.53   &$0.47\pm0.05$  & 10.5\\
RX J0319.8+1845 & w   & 170   & 4.2 & 0.190    & 0.32   &$0.48\pm0.10$  & 4.3\\
RX J0319.8+1845 & on  & 170   & 6.2 & 0.190    & 0.32   &$0.14\pm0.10$  & 4.3\\
1ES 0323+022    & w   & 190   & 8.7 & 0.147    & 0.22   &$1.82\pm0.19$  & 7.8\\
1ES 0414+009    & w   & 190   & 7.7 & 0.287    & 0.57   & \nodata       & 12.1\\
1ES 0806+524    & w   & 190   & 7.6 & 0.138    & 0.21   &$8.00\pm0.23$  & 11.9 \\
1ES 0927+500    & w   & 170   & 7.1 & 0.188    & 0.31   & \nodata       & 9.7\\
1ES 1011+496    & w   & 170   & 10.9/6.5\tablenotemark{f}& 0.212 & 0.37&$11.49\pm0.13$& 5.2\\
1ES 1218+304\tablenotemark{g}&on&120&10.1&0.182& 0.29   &$6.13\pm0.13$  & 21.2\\
1ES 1218+304    & w   & 190   & 7.7 & 0.182    & 0.29   &$4.99\pm0.11$  & 21.2\\
RX J1417.9+2543 & on  & 140   & 2.5 & 0.237    & 0.43   &$2.11\pm0.29$  & 8.7\\
1ES 1426+428    & on  & 140   & 5.5 & 0.129    & 0.19   &$1.87\pm0.15$  & 18.5\\
RX J1725.0+1152 & on  & 190   & 6.3 & $>0.17$  & 0.27\tablenotemark{h}&$13.27\pm0.09$& 8.7\\
1ES 2344+514\tablenotemark{i}&w&180&11.5&0.044 & 0.05   &$3.37\pm0.25$  & 12.0\\
\enddata
\tablenotetext{a}{Peak response energy for a power law spectrum with index -3.0}
\tablenotetext{b}{$\nu F_{\nu}(200\,\rm{GeV})$ in units of $10^{-12}\,\rm{erg}\,\rm{cm}^{-2}\,\rm{s}^{-1}$}
\tablenotetext{c}{optical depth $\tau (z)$ at 200\,GeV}
\tablenotetext{d}{$\nu F_{\nu}(640\,\rm{nm})$  in units of $10^{-12}\,\rm{erg}\,\rm{cm}^{-2}\,\rm{s}^{-1}$ (Host galaxy substracted)}
\tablenotetext{e}{$\nu F_{\nu}(1\,\rm{keV})$ in units of $10^{-12}\,\rm{erg}\,\rm{cm}^{-2}\,\rm{s}^{-1}$ taken from \cite{don}}
\tablenotetext{f}{Flux under the assumption of a detection}
\tablenotetext{g}{Values from \cite{alb2006b}}
\tablenotetext{h}{For a redshift of 0.17}
\tablenotetext{i}{Values from \cite{alb2007b}}
\end{deluxetable}

\section{Discussion}
\subsection{Gamma ray emitting HBLs?}
One may ask whether the gamma ray emitting HBLs can be distinguished from other HBLs based on their spectral energy distributions.  
Finding the answer is hampered by a number of problems.  The variable peak frequencies, which cannot be easily detected in fixed
energy bands, the gamma ray attenuation due to pair production in the metagalactic radiation field, and the flux variability.
From \cite{don}, the amplitude of the flux variability at 1\,keV amounts to a factor of six for the sources with multiple entries in the catalogue. Similar or larger amplitudes can be expected at VHE.  However, our sample is not triggered by flux variability,
the duty cycle of flares generally seems to be rather low,
and the observed fluxes or flux upper limits may therefore be characteristic of the quiescent average fluxes.

\begin{figure}
\plotone{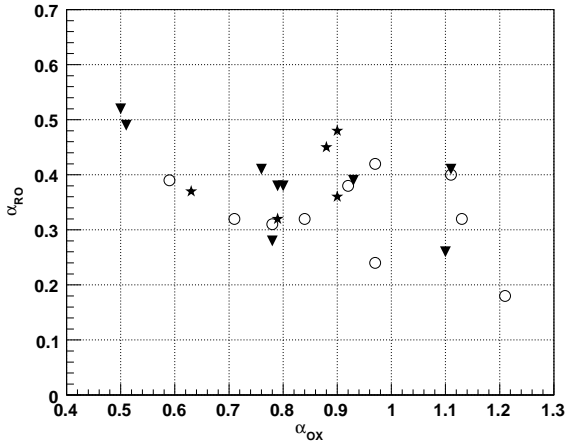} 
\caption{The broad-band spectral index $\alpha_{\rm{RO}}$ vs.\ $\alpha_{\rm{OX}}$. The filled symbols mark the spectral indices of the sources which belong to the sample described in Sect.\,2. They are further divided into detected (stars) and non-detected sources (triangles). The open circles represent all other HBLs detected at VHE $\gamma$-rays so far. 
\label{alpha-ro-ox}}
\end{figure}

Figure\,\ref{alpha-ro-ox} shows the broad-band spectral index $\alpha_{\rm{RO}}$ vs.\ $\alpha_{\rm{OX}}$ for all 24 HBLs as described in the previous
section.
The distribution is quite homogeneous. As the data are not simultaneously taken, the uncertainties due to flux variations have to be taken into account. In the case of $\alpha_{\rm{OX}}$ a flux variability of a factor six at 1\,keV coresponds to a change in the spectral index of 0.29 (if the optical flux remains the same). This is still below the difference of 0.6 between the lowest and highest values of $\alpha_{\rm{OX}}$ for the detected VHE sources. 
Thus there are significant differences in the SED among the HBLs studied here.
As the variability in the radio and optical band for HBLs is lower than at X-rays or VHE $\gamma$-rays and the change of the spectral index $\alpha_{\rm{RO}}$ is smaller for different flux ratios, the variation of $\alpha_{\rm{RO}}$ is much lower than for $\alpha_{\rm{OX}}$. However, it is not possible to distinguish between sources detected at VHE and non-detected ones based on spectral indices.
 
\subsection{Gamma-to-X-ray spectral index}
\begin{figure}[!h]
\plotone{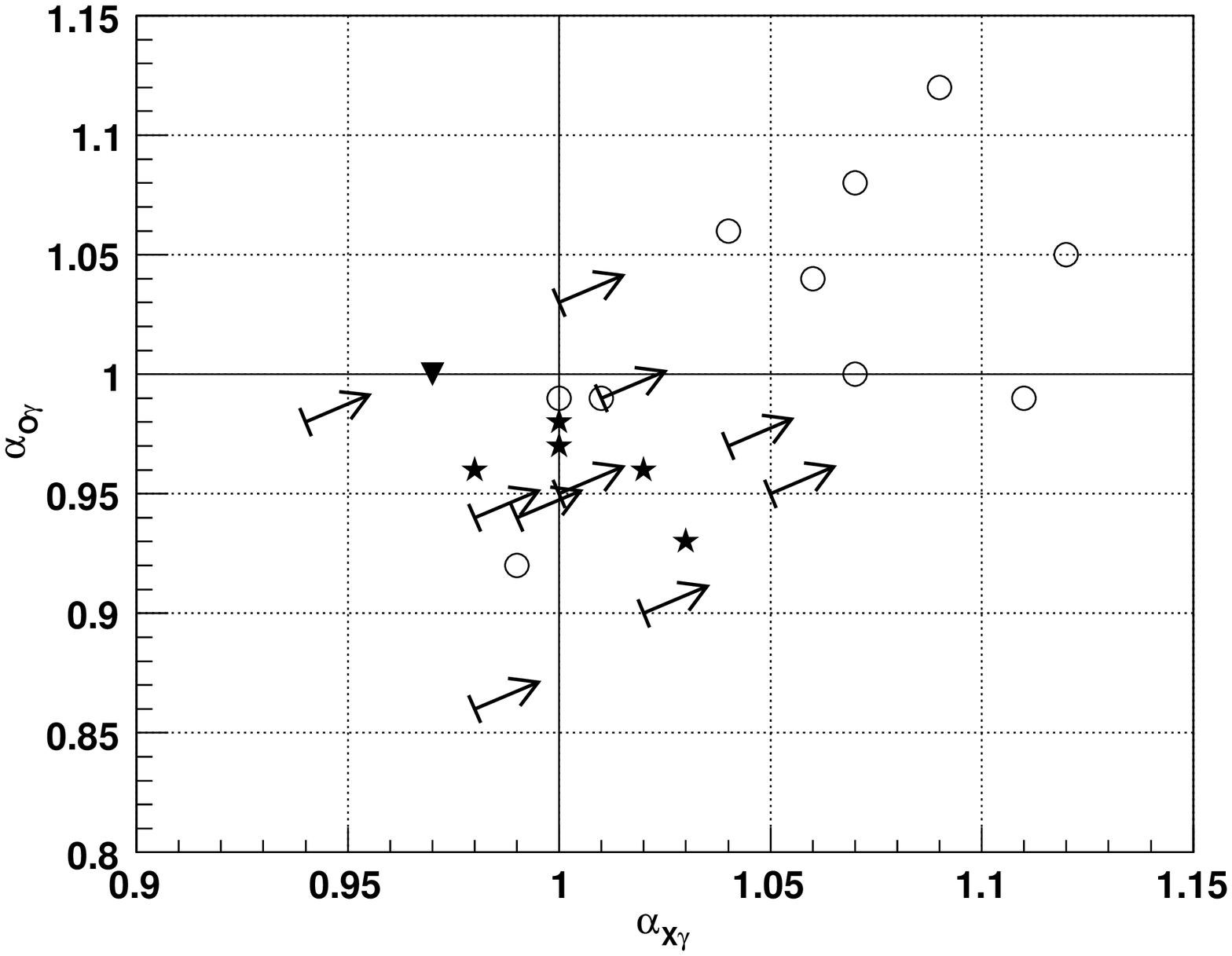} 
\caption{The broad-band spectral index $\alpha_{\rm{O}\gamma}$ vs.\ $\alpha_{\rm{X}\gamma}$. The arrows mark the upper limits for the
spectral indices, as calculated in this work, whereas the stars indicate the spectral indices of the detected sources that belong to the sample described in Sect.\,2. The open circles mark all are other HBL objects detected a VHE $\gamma$-rays so far. 
\label{alpha-og-xg}}
\end{figure}

To unveil the physical state of the emitting plasma, we seek to find the 
characteristic ratio of the two peaks in the spectral energy distribution of HBLs which is related to the ratio of photon and magnetic
field energy densities.
Figure\,\ref{alpha-og-xg} shows the broad-band spectral index $\alpha_{\rm{O}\gamma}$ vs.\ $\alpha_{\rm{X}\gamma}$. Both indices are distributed in a narrow band around unity for all detected sources, where $\alpha = 1$ represents the case that the energy output in both frequency bands is the same. The constraints from the u.l.s on the $\gamma$-ray flux can not exclude the region which is spanned by the detected TeV sources. In the framework of SSC models, the optical and X-ray band belong to synchrotron emission of highly relativistic electrons (first peak in the SED), while the VHE $\gamma$-rays are produced by inverse Compton scattering (second peak in the SED). If the magnetic energy density $u_{\rm{B}}$ is equal to the photon energy density $u_{\rm{ph}}$, the energy output for the synchrotron and the Inverse Compton emission as well as the peak luminosities are the same. In case of HBLs, the peak frequency of the synchrotron emission is always at higher frequencies than the optical band and below 1\,keV for most of them (except the extreme blazars with a hard spectrum at 1\,keV). At VHE energies, most of the detected sources show a soft spectrum above $\sim$\,200\,GeV, indicating observed peak energies below 100\,GeV. Due to absorption of VHE $\gamma$-rays in the MRF, the intrinsic peak energies could also reach several TeV for sources located at higher redshifts. In that context, the scattering of $\alpha_{\rm{O}\gamma}$ and $\alpha_{\rm{X}\gamma}$ around unity could be explained by a continuous distribution of peak frequencies for HBLs, measured with fixed bandwidth.
    
\begin{figure}[!h]
\plotone{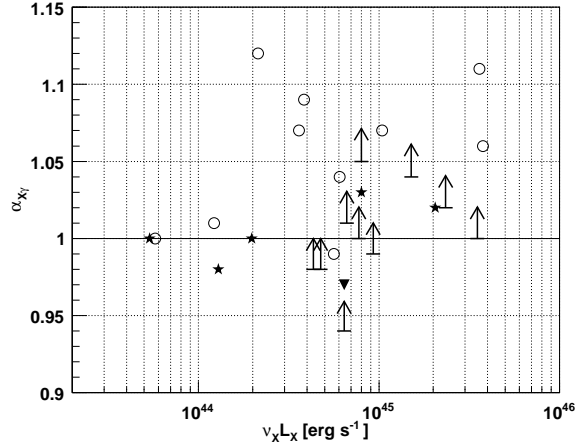} 
\caption{The broad-band spectral index $\alpha_{\rm{X}\gamma}$ vs.\ the X-ray luminosity $\nu_{\rm{X}} L_{\rm{X}}$. The arrows mark the upper limits for the spectral indices, as calculated in this work, while the stars indicate the values for the detected sources that belong to the sample described in Sect.\,2. The open circles mark all other HBL objects detected at VHE $\gamma$-rays so far.
\label{alphaXG-LX}}
\end{figure}

Figure\,\ref{alphaXG-LX} shows the broad-band spectral index $\alpha_{X\gamma}$ vs.\ X-ray luminosity $\nu_{\rm{X}}L_{\rm{X}}$. The average energy output at 1\,keV never exceeds significantly the one at 200\,GeV ($\alpha_{X\gamma} > 0.97$). For half of the detected sources, the energy output in these bands is almost the same ($\alpha_{X\gamma} \cong 1$), while for the other ones the energy output at 200\,GeV is significantly lower. There is a tendency that this effect shows up at higher X-ray luminosities. 
For four of the ten u.l., $\alpha_{X\gamma} = 1$ can not be excluded, whereas for the other six sources $\alpha_{X\gamma}$ has to be larger than unity. For the non-detected sources of the sample, further observations with longer exposures are needed to reach the $\alpha_{X\gamma} = 1.12$--line (corresponding to a nine times lower output at 200\,GeV compared to 1\,keV), which includes all HBLs detected at VHE so far. Note that if $\alpha_{X\gamma} = 1$ for the peak frequencies is valid, the tendency of increasing $\alpha_{X\gamma}$ values with increasing luminosities at 1\,keV could be interpreted as a shift of the inverse Compton peak to lower values.
Similar studies aiming at a characterization of VHE blazars are also pursued by \citet{wagner07}.

\subsection{Constraints on the luminosity function of $\gamma$-ray emitting HBLs}
\begin{figure}[!h]
\plotone{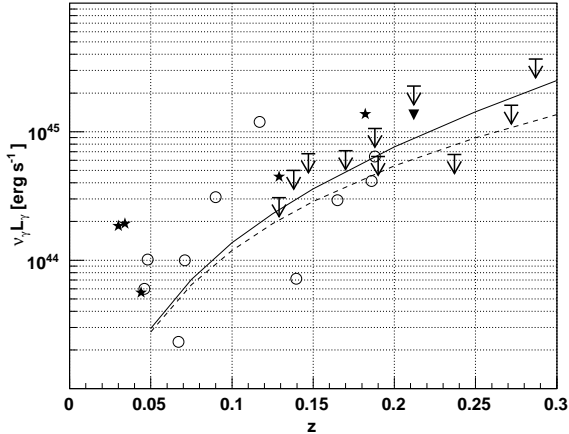} 
\caption{The $\gamma$-ray luminosity $\nu_{\gamma} L_{\gamma}$ vs.\ redshift. The arrows mark the u.l.s calculated in this work, whereas the stars indicate the detected sources that belong to the sample described in Sect.\,2. The triangle marks 1ES 1011+496 if interpreted as a detection. The open circles mark all other HBL objects detected at VHE $\gamma$-rays so far. The dashed line indicates a flux of $4.8\times10^{-12}\,\rm{erg}\,\rm{cm}^{-2}\,\rm{s}^{-1}$ corresponding to $2\,\mu\rm{Jy}$ at 1\,keV. The solid line corresponds to
the same flux taking into account $\gamma$-ray attenuation at 200\,GeV. \label{LG-Z}}
\end{figure}

Figure\,\ref{LG-Z} shows the luminosity $\nu_{\gamma}L_{\gamma}$ at 200\,GeV vs.\ the redshift. All detected sources are above or within the line that marks the corresponding luminosity to a flux of $2\,\mu\rm{Jy}$ at 1\,keV. The absorption of $\gamma$-rays by the MRF increases with redshift, so that
at a redshift of $z=0.3$, the emitted luminosity becomes twice larger than the measured one.

The luminosity function at $\gamma$-ray energies of HBLs is poorly known, since there has not been a complete survey and the number of detected sources is still rather low. Nevertheless, we can try to constrain the VHE luminosity function from our observations.
To this end we derive upper limits on the cumulative omnidirectional flux at 200\,GeV from X-ray bright HBLs below $z=0.3$, noting that GLAST will measure the diffuse extragalactic background up to 200\,GeV. The selection criterion for the declination of the sample corresponds the a patch of the sky with a size of 5.55 sr (or 44\% of the sky). The sum over all 14 sources of the sample divided by the 5.55\,sr patch results in an upper limit on the total intensity at 200 GeV of $I_{\rm VHE}(200~{\rm GeV}) = \varepsilon \cdot 2.76 \times 10^{-9} \rm GeV\,cm^{-2}\,sr^{-1}\,s^{-1}$, where $\varepsilon$ accounts for the incompleteness of the sample. Assuming an isotropic distribution of HBLs, $\varepsilon$ should be larger than unity. We know about two sources, not included in this calculation (1ES 1727+502 and 1ES 0229+200). There are also 5 HBLs from \cite{don} that fullfill the criteria of declination and X-ray flux, but are located at higher redshifts. Including these sources the sample would increase to 21 sources. With respect to the $\sim\,50$\% sky coverage of the Einstein Slew Survey, where most of the sources belong to, $\varepsilon=3$ seems to be a reasonable assumption (factor 1.5 to take the "known" sources into account and a factor 2 for the assumed sky coverage of 50\% for the extended sample of 21 sources). This would lead to an upper limit for the total intensity at 200\,GeV of $I_{\rm VHE}({\rm 200~GeV})= 8.3\times 10^{-8}\rm GeV\,cm^{-2}\,sr^{-1}\,s^{-1}$. Not that this u.l.\ is already conservative as also sources are included which are not in the Einstein Slew Survey (would result in a higher sky coverage) and the u.l.\ is dominated by the two brightest sources Mrk 501 and Mrk 421 (which would mean that three times more sources would not necessarily lead to a three times higher flux). Recently \cite{kneiske07a} showed that HBLs could contribute up to 30\% of the extragalactic background radiation at GeV energies, when including cascade emission from sources at higher redshifts. The luminosity function used for their calculation was derived from the X-ray luminosity function \citep{beckmann2003} assuming the same luminosity above 300\,GeV as from 0.5\,keV to 2\,keV. For the HBL contribution of faint point sources at 200\,GeV, they obtain (somewhat model-dependent) 
$I_{\rm KM}^{\rm (point)}({\rm 200\,GeV})= (4-6) \times 10^{-8}\,\rm GeV\, cm^{-2}\,sr^{-1}\,s^{-1}$ which is within the upper limit obtained here.  
For the total intensity, including the diffuse component due to electromagnetic cascading, their result is
$I_{\rm KM}^{\rm (diffuse)}({\rm 200\,GeV})= 1.0 \times 10^{-7}\,\rm GeV\, cm^{-2}\,sr^{-1}\,s^{-1}$.  This estimate is based on assuming
that the emitted VHE spectra generally have peaks at energies well in excess of 200\,GeV. 

\section{Conclusions}
Searching for VHE $\gamma$-ray emission from a sample of twelve X-ray bright HBL objects, 1ES 1218+30.4 at a redshift of $z=0.182$ was discovered
for the first time at VHE \citep{alb2006b}.  
The already established VHE source 1ES 2344+514 has been detected with high significance\citep{alb2007b}, albeit in a state of low activity.
For ten sources no significant signal was seen, resulting in upper limits on their integral flux above $\sim$\,200\,GeV between 2.3\% and 8.6\% of the
Crab Nebula flux on a 99\% confidence level. There is a hint for a signal from 1ES 1011+496 on a $3.5\,\sigma$ level, which has been confirmed
as a source of VHE $\gamma$-rays by a second MAGIC observation campaign, triggered by a high optical state \citep{alb2007e}.

With fixed-schedule observations a bias to flaring emission states was avoided, tacitly assuming that the duty cycle of the flares is short when compared with the exposure time. As shown in the case of 1ES 2344+514, quiescence does not necessarily preclude the sources from being detected.
As a matter of fact, the upper limits obtained for the other sources still lie in a region of parameter space bracketed by the detected
sources, and correspond to a VHE energy flux on the level of the X-ray energy flux with an indication of an increasing $\alpha_{\rm{X}\gamma}$ with increasing X-ray luminosity.  It thus seems to be a question of time that more sensitive telescopes, and in particular those with a lowered energy threshold such as MAGIC-II, potentially capturing shifting peaks, will eventually lead to a detection of all known bright HBLs. 
It will be important to obtain a complete catalogue of HBLs from the planned eROSITA and GLAST all-sky surveys 
to study them on a much larger statistical
basis in the future.

The detected sources deviate in no apparent pattern from the so-far non-detected sources.  A spectral shape with two equal-height bumps is expected in Synchrotron Self Compton models for the case of balanced energy densities of photons and the magnetic field. The dispersion in the distribution of the X-to-$\gamma$-ray luminosity ratio (the lowest X-to-$\gamma$-ray luminosity ratio of a detected HBL is 1/9)
would then reflect variations of the peak position w.r.t.\ the observed band or the effect of flux variability (e.g., high state X-ray emission vs.
quiescent VHE emission).

\acknowledgements
We would like to thank the IAC for the excellent working
conditions at the Observatorio del Roque de los Muchachos in La Palma.
The support of the German BMBF and MPG, the Italian INFN and the
Spanish CICYT is gratefully acknowledged. This work was also
supported by ETH Research Grant TH~34/04~3 and the Polish MNiI Grant
1P03D01028.


\begin{thebibliography}{}
%
\bibitem[Aharonian (2000)]{aha2000}
Aharonian, F.~A. 2000, New Astronomy, 5, 377
\bibitem[Aharonian et al.(2002)]{aha02}
Aharonian, F., Akhperjaan, A., Barrio, J., et al.\ 2002, A\&A, 384, L23
%
\bibitem[Aharonian et al.(2003)]{aha03}
Aharonian, F., Akhperjaan, A., Beilicke, M., et al.\ 2003, A\&A, 403, 523
%
\bibitem[Aharonian et al.(2005a)]{aha2005a}
Aharonian, F., Akhperjanian, A., Aye, K.-M.\ et al.\ 2005a, A\&A, 430, 865
%
\bibitem[Aharonian et al.(2005b)]{aha2005b}
Aharonian, F., Akhperjanian, A., Aye, K.-M.\ et al.\ 2005b, A\&A, 436, L17
%
\bibitem[Aharonian et al.(2006a)]{aha2006a}
Aharonian, F., Akhperjanian, A. G., Bazer-Bachi, A. R. et al.\ 2006, Nature, 440, 1018
%
\bibitem[Aharonian et al.(2006b)]{aha2006b}
Aharonian, F., Akhperjanian, A. G., Bazer-Bachi, A. R. et al.\ 2006, Science, 314, 1424 
%
\bibitem[Aharonian et al.(2007a)]{aha2007a}
Aharonian, F., Akhperjanian, A. G., Bazer-Bachi, A. R. et al.\ 2007, \apjl, 664, L71 
%
\bibitem[Aharonian et al.(2007b)]{aha2007b}
Aharonian, F., Akhperjanian, A. G., Barres de Almeida, U. et al.\ 2007, accepted by A\&A, arXiv:0708.3021v1 [astro-ph] 
%
\bibitem[Aharonian et al.(2007c)]{aha2007c}
Aharonian, F., Akhperjanian, A. G., Barres de Almeida, U. et al.\ 2007, accepted for publication in A\&A Letters, arXiv:0709.4584v1 [astro-ph] 
%
\bibitem[Albert et al.(2006a)]{alb2006a}
Albert, J., Aliu, E., Anderhub, H.,et al. 2006a \apj, 639, 761
\bibitem[Albert et al.(2006b)]{alb2006b}
Albert, J., Aliu, E., Anderhub, H.,et al. 2006b \apjl, 642, L119
%
\bibitem[Albert et al.(2006c)]{alb2006c}
Albert, J., Aliu, E., Anderhub, H.,et al. 2006c \apjl, 648, L105
%
\bibitem[Albert et al.(2007a)]{alb2007a}
Albert, J., Aliu, E., Anderhub, H.,et al. 2007a, \apj, 654, L119 
%
\bibitem[Albert et al.(2007b)]{alb2007b}
Albert, J., Aliu, E., Anderhub, H.,et al. 2007c, \apj, 662, 892
%
\bibitem[Albert et al.(2007c)]{alb2007c}
Albert, J., Aliu, E., Anderhub, H.,et al. 2007b, \apj, 663, 125
%
\bibitem[Albert et al.(2007d)]{alb2007d}
Albert, J., Aliu, E., Anderhub, H.,et al. 2007e, \apjl, 666, L17
%
\bibitem[Albert et al.(2007e)]{alb2007e}
Albert, J., Aliu, E., Anderhub, H.,et al. 2007g, \apjl, 667, L21
%
\bibitem[Albert et al.(2007f)]{alb2007f}
Albert, J., Aliu, E., Anderhub, H.,et al. 2007d, \apj, 669, 862
%
\bibitem[Albert et al.(2007g)]{alb2007g}
Albert, J., Aliu, E., Anderhub, H.,et al. 2007f, \apj, in press, arXiv:0705.3244v1 [astro-ph]
%
%
\bibitem[Baixeras et al.(2004)]{MAGIC-commissioning}
Baixeras, C., Bastieri, D., Bigongiari, C., et al. 2004,
Nucl.\ Instrum.\ Meth., A518, 188
%
\bibitem[Beckmann et al.(2003)]{beckmann2003}
Beckmann, V., Engels, D., Bade, N.\ \& Wucknitz, O. 2003, A\&A, 401, 927
\bibitem[Bersanelli et al.(1992)]{bersanelli}
Bersanelli, M., Bouchet, P., Falomo, R.\ \& Tanzi, E.~G.\ 1992, AJ, 104, 28
%
\bibitem[Biretta et al.(1991)]{biretta1991}
Biretta, J.~A., Stern, C.~P.\ \& Harris, D.~E. 1991, AJ, 101, 1632
%
\bibitem[Biretta et al.(2002)]{biretta2002}
Biretta, J.~A., Junor, W.\ \& Livio, M. 2002, NewAR, 46, 239
%
\bibitem[Bretz(2005a)]{mars} 
Bretz, T. 2005a, in AIP Conf. Proc. 745, 730
%
\bibitem[Bretz(2005b)]{bre2} 
Bretz, T. 2005b, Proc.\ of the 29th ICRC, Pune, India
%
\bibitem[Cortina et al.(2005)]{CortinaICRC}
Cortina, J., Armada, A., Biland, A., et al. 2005,
Proc.\ of the 29th ICRC, Pune, India, astro-ph/0508274
%
\bibitem[Costamante \& Ghisellini(2002)]{CostGhis}
Costamante, L. \& Ghisellini, G. 2002, A\&A, 384, 56
%
\bibitem[Dermer \& Schlickeiser(1993)]{der}
Dermer, C.~D. \& Schlickeiser, R. 1993, ApJ, 416, 458
%
\bibitem[Donato et al.(2001)]{don}
Donato, D., Ghisellini, G., Tagliaferri, G. \& Fossati, G. 2001, A\&A,
375, 739
%
\bibitem[Elvis et al.(1992)]{elvis}
Elvis, M., Plummer, D., Schachter, J.\ \& Fabbiano, G.\ 1992, ApJS, 80, 257
%
\bibitem[Fazio \& Stecker(1979)]{fazio}
Fazio, G.G., Stecker, F.W. 1979, Nature, 226, 135
%
\bibitem[Gaug et al.(2005)]{gaug}
Gaug, M., Bartko, H., Cortina, J. \& Rico, J. 2005, Proc.\ of the 29th
ICRC, Pune, India, astro-ph/0508274
%
\bibitem[Ghisellini et al.(1993)]{ghisellini1993}
Ghisellini, G., Padovani, P., Celotti, A.\ \& Maraschi, L. 1993, ApJ, 407, 65
%
\bibitem[Giommi \& Padovani(1994)]{giommi1994}
Giommi, P., Padovani, P. 1994, \mnras, 268, L51
%
\bibitem[Goebel et al.(2007)]{goebel2007}
Goebel, F., Bartko, H., Carmona, E.\ et al. 2007, to appear in the Proc. of the 30th Internat.\ Cosmic Ray Conf., Merida, Mexico, arXiv:0709.2363v1 [astro-ph]

\bibitem[Hauser \& Dwek(2001)]{dwek}
Hauser, M.G. \& Dwek, E. 2001, ARA\&A, 39, 249
%
\bibitem[Heck et al.(1998)]{cor}
Heck, D., Knapp, J., Capdevielle, J. N., Schatz, G. \& Thouw, T. 1998,
Report FZKA 6019, Forschungszentrum Karlsruhe;
http://www-ik.fzk.de/\symbol{126}heck/corsika/physics\_description/ corsika\_physics.html 
%
\bibitem[Hillas(1985)]{Hillas_parameters}
Hillas, A. M. 1985, Proc.\ of the 19th ICRC, La Jolla, 3, 445
%
\bibitem[Horan et al.(2002)]{hor}
Horan, D., Badran, H. M., Bond, I. H., et al.\ 2002, ApJ, 571, 753
%
\bibitem[Kneiske et al.(2004)]{kneiske04}
Kneiske, T. M., Bretz, T., Mannheim, K. \& Hartmann, D.H. 2004, A\&A,
413, 807
%
\bibitem[Kneiske \& Mannheim(2007)]{kneiske07a}
Kneiske, T. M.\ \& Mannheim, K.\ 2007, accepeted by A\&A, arXiv:0705.3778v1
%
\bibitem[Landt(2003)]{landt}
Landt, H. 2003, phd thesis, Hamburg
%
\bibitem[Lessard et al.(2001)]{les}
Lessard, R.W., Buckley J. H., Connaughton, V. \& Le Bohec, S. 2001,
Astroparticle Physics, 15, 1
%
\bibitem[Li \& Ma(1983)]{lima}
Li, T.\ \& Ma, Y.\ 1983, ApJ, 272, 317
%
\bibitem[Majumdar et al.(2005)]{maj}
Majumdar, P., Moralejo, A., Bigongiari, C., Blanch, O.\ \& Sobczynska,
D.\ 2005, Proc.\ of the 29th ICRC, Pune, India, astro-ph/0508274
%
\bibitem[Mannheim(1993)]{man}
Mannheim, K.\ 1993, A\&A, 269, 67
%
\bibitem[Maraschi et al.(1992)]{mar}
Maraschi, L., Ghisellini, G., \& Celotti, A. 1992, ApJ, 397, L5
%
\bibitem[Muecke \& Protheroe(2001)]{pro}
Muecke, A. \& Protheroe, R. J. 2001, Astropart. Phys. 15, 121
%
\bibitem[Nilsson et al.(2007)]{nilsson}
Nilsson, K., Pasanen, M., Takalo, L., et al. 2007, A\&A, 472, 199
%
\bibitem[Petry et al.(2002)]{pet}
Petry, D., Bond, I.H., Bradbury, S. M. 2002, ApJ, 580, 104
%
\bibitem[Riegel(2005)]{dyn_cuts}
Riegel, B. Bretz, T. et al. 2005, Proc.\ of the 29th ICRC, Pune, India 
%
\bibitem[Rolke et al.(2005)]{rol}
Rolke, W., Lopez, A., Conrad, J. \& James, F. 2005, Nucl.Instrum.Meth., A551, 493
%
\bibitem[Sbarufatti et al.(2006)]{sbaru2006}
Sbarufatti, B, Treves, A., Falomo, R. et al. 2006, AJ, 132, 1
%
\bibitem[Schlegel et al.(1998)]{schlegel}
Schlegel, D.~J., Finkbeiner, D.~P.\ \& Davis, M.\ 1998, ApJ, 500, 525S
%
\bibitem[Sikora et al.(1994)]{sikora94}
Sikora, M.~, Begelmann, M.~C.~ \& Rees, M.~J.~ 1994, ApJ, 421, 153
%
\bibitem[Stecker \& Scully(2007)]{stecker2007}
Stecker, F.~W.\ \& Sully, S.~T. 2007, A\&A in press, arXiv:0710.2252v1 [astro-ph]
%
\bibitem[Strong et al.(2004)]{strong}
Strong, A.~W., Moskalenko, I.~V.\ \& Reimer, O.\ 2004, ApJ, 613, 956
%
\bibitem[Superina et al.(2007)]{superina07}
Superina, G., Benbow, W., Boutelier, T., Dubus, G.\ \& Giebels, B. 2007, to appear in the Proc. of the 30th Internat.\ Cosmic Ray Conf., Merida, Mexico, arXiv:0710.4057
%
\bibitem[Teshima et al.(2007)]{teshima2007}
Teshima, M., Prandini, E., Bock, R.\ et al. 2007, to appear in the Proc. of the 30th Internat.\ Cosmic Ray Conf., Merida, Mexico, arXiv:0709.1475v1 [astro-ph]
%
\bibitem[Urry \& Padovani(1995)]{urrypad95}
Urry, C.~M.~ \& Padovani, P. 1995, PASP, 107, 803
%
\bibitem[Wagner (2007)]{wagner07}
Wagner, R.~M. 2007, MNRAS accepted, arXiv:0711.3025 [astro-ph]
\end{thebibliography}
\end{document}